\documentclass{article}
\usepackage[utf8]{inputenc}
\usepackage{lineno,hyperref}
\usepackage{spconf,amsmath,graphicx,bm,amssymb}
\graphicspath{ {images/} }
\usepackage{textcomp}
\usepackage{multirow}
\usepackage[table]{xcolor}
\usepackage{booktabs}
\newcommand{\cg}{\cellcolor{gray!30}}
\newcommand{\rowgroup}[1]{\hspace{1em}#1}

\title{The Intelligent Voice 2016 Speaker Recognition System}
\name{Abbas Khosravani, Cornelius Glackin, Nazim Dugan, Gérard Chollet, Nigel Cannings}
\address{Intelligent Voice Limited, St Clare House, 30-33 Minories, EC3N 1BP, London, UK}
\begin{document}
%\ninept
%
\maketitle
\begin{abstract}
This paper presents the Intelligent Voice (IV) system submitted to the NIST 2016 Speaker Recognition Evaluation (SRE). The primary emphasis of SRE this year was on developing speaker recognition technology which is robust for novel languages that are much more heterogeneous than those used in the current state-of-the-art, using significantly less training data, that does not contain meta-data from those languages. The system is based on the state-of-the-art \emph{i}-vector/PLDA which is developed on the fixed training condition, and the results are reported on the protocol defined on the development set of the challenge.

\end{abstract}
\section{Introduction}
\label{sec:intro}
Compared to previous years, the 2016 NIST speaker recognition evaluation (SRE) marked a major shift from English towards Austronesian and Chinese languages. The task like previous years is to perform speaker detection with the focus on telephone speech data recorded over a variety of handset types. The main challenges introduced in this evaluation are duration and language variability. The potential variation of languages addressed in this evaluation, recording environment, and variability of test segments duration influenced the design of our system. Our goal was to utilize recent advances in language normalization, domain adaptation, speech activity detection and session compensation techniques to mitigate the adverse bias introduced in this year's evaluation.

Over recent years, the \emph{i}-vector representation of speech segments has been widely used by state-of-the-art speaker recognition systems \cite{dehak2011front}. The speaker recognition technology based on \emph{i}-vectors currently dominates the research field due to its performance, low computational cost and the compatibility of \emph{i}-vectors with machine learning techniques. This dominance is reflected by the recent NIST \emph{i}-vector machine learning challenge \cite{greenberg2014nist} which was designed to find the most promising algorithmic approaches to speaker recognition specifically on the basis of \emph{i}-vectors  \cite{khosravani2014linearly,novoselov2014stc,vesnicer2014incorporating,khoury2014hierarchical}. The outstanding ability of DNN for frame alignment which has achieved remarkable performance in text-independent speaker recognition for English data \cite{lei2014novel,kenny2014deep}, failed to provide even comparable recognition performance to the traditional GMM. Therefore, we concentrated on the cepstral based GMM/\emph{i}-vector system.

We outline in this paper the Intelligent Voice system, techniques and results obtained on the SRE 2016 development set that will mirror the evaluation condition as well as the timing report. Section \ref{sec:data} describes the data used for the system training. The front-end and back-end processing of the system are presented in Sections \ref{sec:frontend} and \ref{sec:backend} respectively. In Section \ref{sec:results}, we describe experimental evaluation of the system on the SRE 2016 development set. Finally, we present a timing analysis of the system in Section \ref{sec:timing}. 

\section{Training Condition}
\label{sec:data}
The fixed training condition is used to build our speaker recognition system. Only conversational telephone speech data from datasets released through the linguistic data consortium (LDC) have been used, including NIST SRE 2004-2010 and the Switchboard corpora (Switchboard Cellular Parts I and II, Switchboard2 Phase I,II and III) for different steps of system training. A more detailed description of the data used in the system training is presented in Table \ref{tab:data-nist}. We have also included the unlabelled set of 2472 telephone calls from both minor (Cebuano and Mandarin) and major (Tagalog and Cantonese) languages provided by NIST in the system training. We will indicate when and how we used this set in the training in the following sections.

\begin{table}[ht]
  \renewcommand{\arraystretch}{1}
  \setlength{\tabcolsep}{4pt}
  \vspace{-2mm}
  \caption{\label{tab:data-nist}\it{The description of the data used for training the speaker recognition system.}}
  \vspace{1mm}
  \centerline
  {
    \begin{tabular}{l c c c c c c c c c c c c}
      \toprule
      & & \multirow{2}{*}{\bf \#Langs} & & \multicolumn{2}{c}{\bfseries{\#Spks}} & & \multicolumn{2}{c}{\bfseries{\#Segs}} \\
      \cmidrule{5-6} \cmidrule{8-9} 
       & & & & Male & Female & & Male & Female \\
      \midrule
       English &   & 1 & & 1925 & 2603 & & 19556 & 25835\\
       non-English &   & 34 & & 274 & 489 & & 1428 & 2657 \\
      \bottomrule
    \end{tabular}
  }
\vspace{-2mm}
\end{table}

\section{Front-End Processing}
\label{sec:frontend}
In this section we will provide a description of the main steps in front-end processing of our speaker recognition system including speech activity detection, acoustic and \emph{i}-vector feature extraction. 

\subsection{Speech Activity Detection}
\label{sec:vad}

\begin{figure}[t]
    \includegraphics[width=8cm]{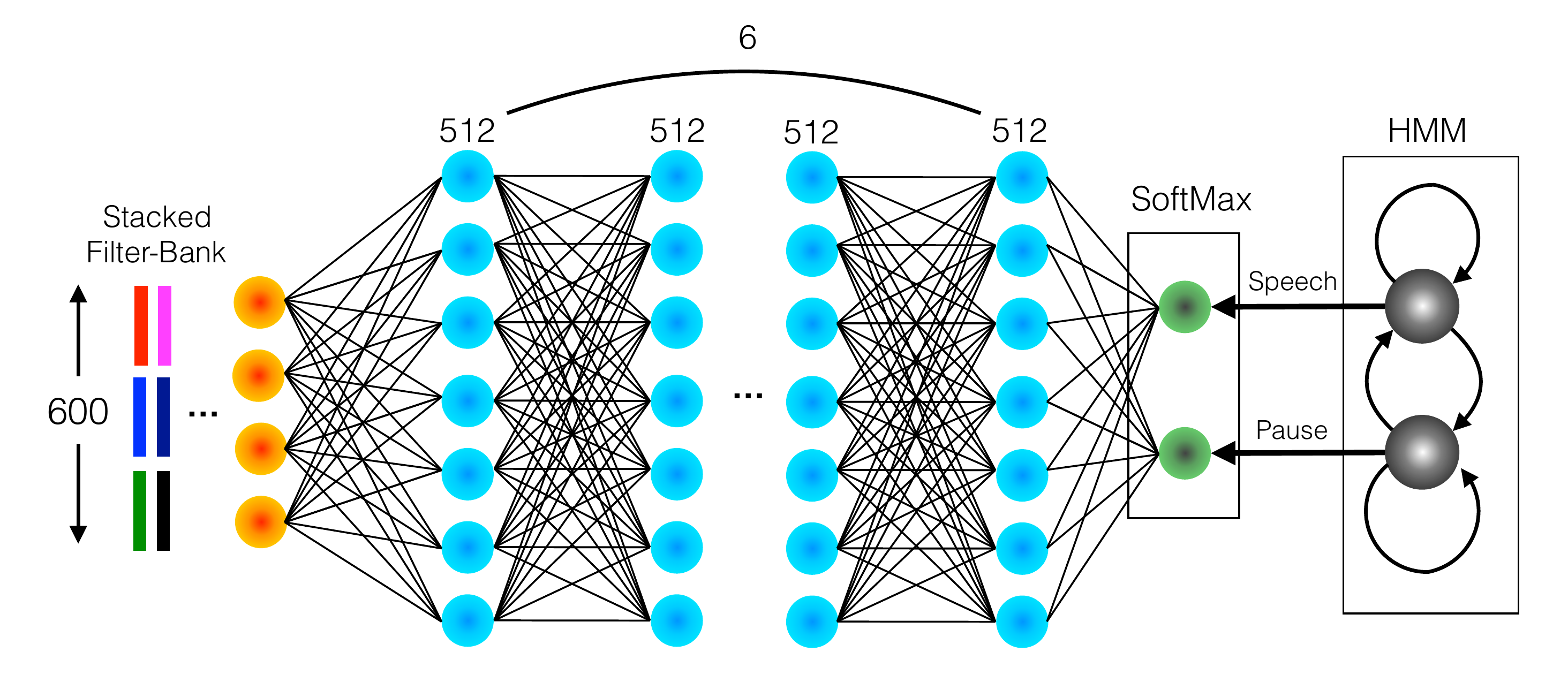}
    \centering
    \caption{{\it The architecture of our DNN-HMM speech activity detection.}}
    \label{fig:DNN-SAD}
\end{figure}

The first stage of any speaker recognition system is to detect the speech content in an audio signal. An accurate speech activity detector (SAD) can improve the speaker recognition performance. Several techniques have been proposed for SAD, including unsupervised methods based on a thresholding signal energy, and supervised methods that train a speech/non-speech classifier such as support vector machines (SVM) \cite{mesgarani2006discrimination} and Gaussian mixture models (GMMs) \cite{ng2012developing}. Hidden markov models (HMMs) \cite{pfau2001multispeaker} have also been successful. Recently, it has been shown that DNN systems achieve impressive improvement in performance especially in low signal to noise ratios (SNRs) \cite{ryant2013speech}. In our work we have utilized a two-class DNN-HMM classifier to perform this task. The DNN-HMM hybrid configuration with cross-entropy as the objective function has been trained with the back-propagation algorithm. The softmax layer produces posterior probabilities for speech and non-speech which were then converted into log-likelihoods. Using 2-state HMMs corresponding to speech and non-speech, frame-wise decisions are made by Viterbi decoding. As input to the network, we fed 40-dimensional filter-bank features along with 7 frames from each side. The network has 6 hidden layers with 512 units each. The architecture of our DNN-HMM SAD is shown in Figure \ref{fig:DNN-SAD}.
Approximately 100 hours of speech data from the Switchboard telephony data with word alignments as ground-truth were used to train our SAD. The DNN training in performed on an NVIDIA TITAN X GPU, using Kaldi software \cite{povey2011kaldi}. Evaluated on 50 hours of telephone speech data from the same database, our DNN-HMM SAD indicated a frame-level miss-classification (speech/non-speech) rate of 5.9\% whereas an energy-based SAD did not perform better than 20\%.

\subsection{Acoustic Features}
\label{sec:feats}
For acoustic features we have experimented with different configurations of cepstral features. We have used 39-dimensional PLP features and 60-dimensional MFCC features (including their first and second order derivatives) as acoustic features. Moreover, our experiments indicated that the combination of these two feature sets performs particularly well in score fusion. Both PLP and MFCC are extracted at 8kHz sample frequency using Kaldi \cite{povey2011kaldi} with 25 and 20 ms frame lengths, respectively, and a 10 ms overlap (other configurations are the same as Kaldi defaults). For each utterance, the features are centered using a short-term (3s window) cepstral mean and variance normalization (ST-CMVN). Finally, we employed our DNN-HMM speech activity detector (SAD) to drop non-speech frames.

\subsection{\emph{i}-Vector Features}
\label{sec: ivector}
Since the introduction of \emph{i}-vectors in \cite{dehak2011front}, the speaker recognition community has seen a significant increase in recognition performance. \emph{i}-Vectors are low-dimensional representations of Baum-Welch statistics obtained with respect to a GMM, referred to as \emph{universal background model} (UBM), in a single subspace which includes all characteristics of speaker and inter-session variability, named \emph{total variability matrix} \cite{dehak2011front}. We trained on each acoustic feature a full covariance, gender-independent UBM model with 2048 Gaussians followed by a 600-dimensional \emph{i}-vector extractor to establish our MFCC- and PLP-based \emph{i}-vector systems. The unlabeled set of development data was used in the training of both the UBM and the \emph{i}-vector extractor. The open-source Kaldi software has been used for all these processing steps \cite{povey2011kaldi}. 

It has been shown that successive acoustic observation vectors tend to be highly correlated. This may be problematic for maximum a posteriori (MAP) estimation of \emph{i}-vectors. To investigating this issue, scaling the zero and first order Baum-Welch statistics before presenting them to the \emph{i}-vector extractor has been proposed. It turns out that a scale factor of 0.33 gives a slight edge, resulting in a better decision cost function \cite{kenny2013plda}. This scaling factor has been performed in training the \emph{i}-vector extractor as well as in the testing.

\section{Back-End Processing}
\label{sec:backend}
This section provides the steps performed in back-end processing of our speaker recognition system.

\subsection{Nearest-neighbor Discriminant Analysis (NDA)}
\label{sec:white}
The nearest-neighbor discriminant analysis is a nonparametric discriminant analysis technique which was proposed in \cite{fukunaga1983nonparametric}, and recently used in speaker recognition \cite{Sadjadi2016}. The nonparametric within- and between-class scatter matrices $\hat{\mathbf{S}}_{w}$ and $\hat{\mathbf{S}}_{b}$, respectively, are computed based on $k$ nearest neighbor sample information. The NDA transform is then formed using eigenvectors of $\hat{\mathbf{S}}_{w}^{-1}\hat{\mathbf{S}}_{b}$. It has been shown that as the number of nearest neighbors $k$ approaches the number of samples in each class, the NDA essentially becomes the LDA projection. Based on the finding in \cite{Sadjadi2016}, NDA outperformed LDA due to the ability in capturing the local structure and boundary information within and across different speakers. We applied a $600\times400$ NDA projection matrix computed using the 10 nearest sample information on centered \emph{i}-vectors. The resulting dimensionality reduced \emph{i}-vectors are then whitened using both the training data and the unlabelled development set.

\subsection{Short-Duration Variability Compensation}
\label{sec:short}
\begin{figure}[t]
    \includegraphics[width=8cm]{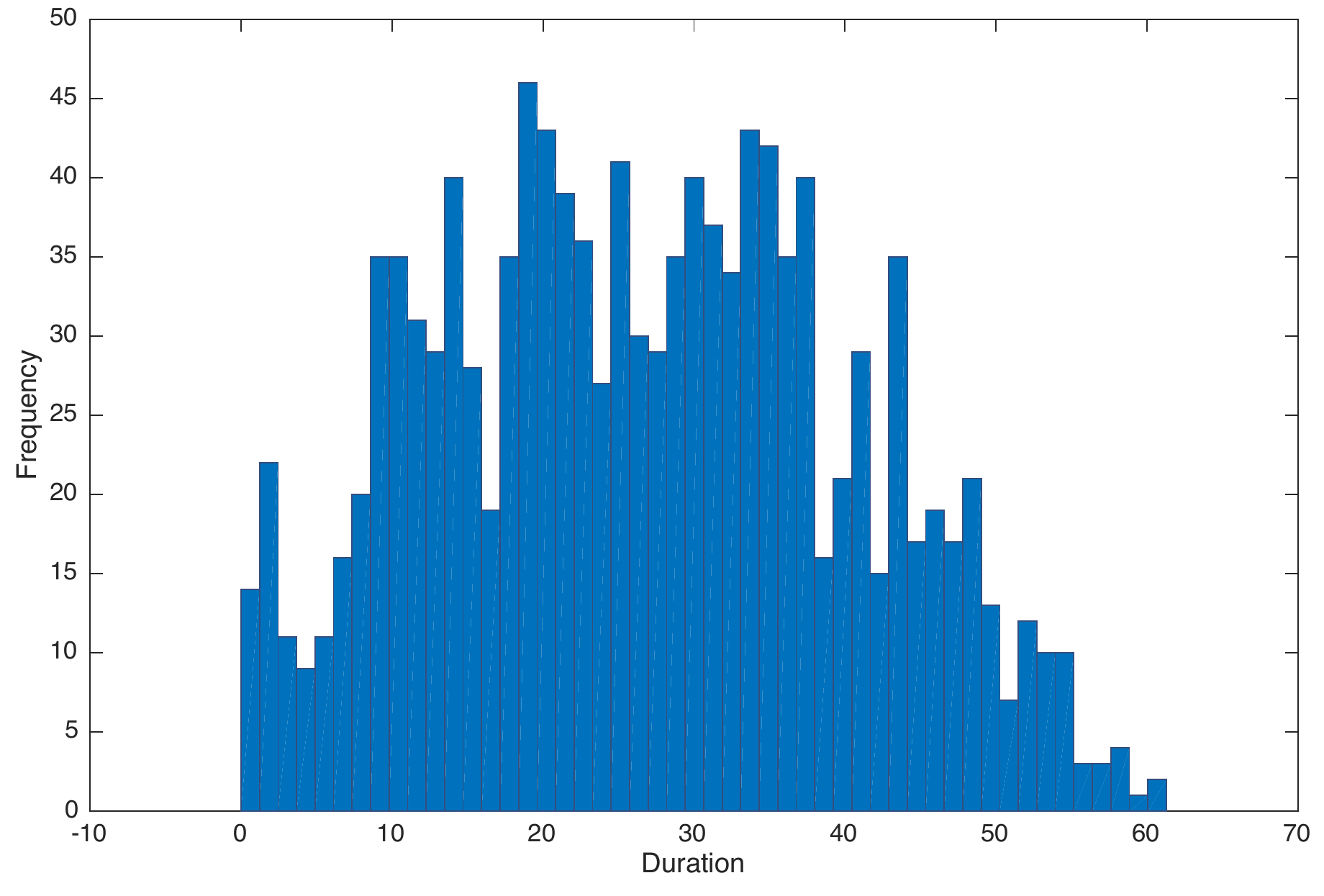}
    \centering
    \caption{{\it The duration of test segments in the development set after dropping non-speech frames.}}
    \label{fig:Duration}
\end{figure}

\begin{figure}[t]
    \includegraphics[width=7cm]{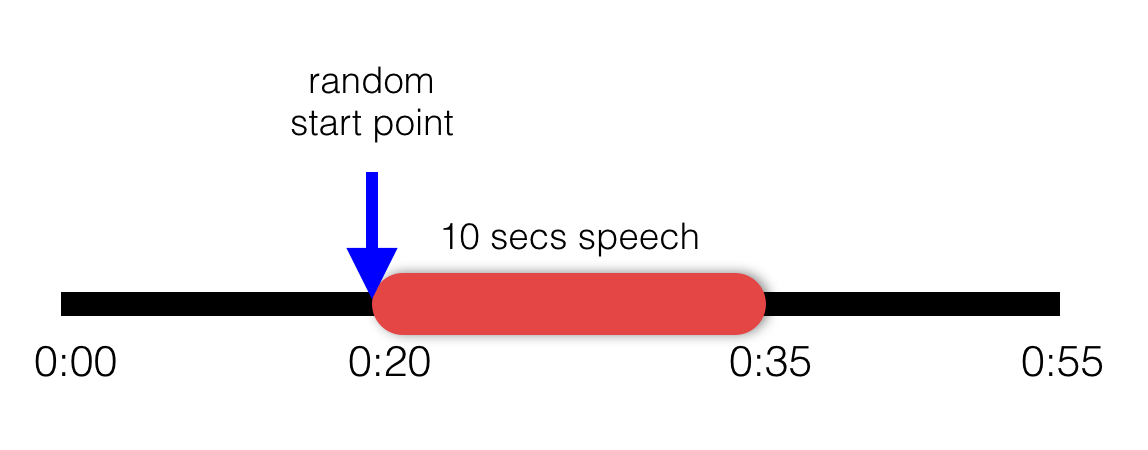}
    \centering
    \caption{{\it Partial excerpt of 10 second speech duration from an audio speech file.}}
    \label{fig:short}
\end{figure}

The enrolment condition of the development set is supposed to provide at least 60 seconds of speech data for each target speaker. Nevertheless, our SAD indicates that the speech content is as low as 26 seconds in some cases. The test segments duration which ranges from 9 to 60 seconds of speech material can result in poor performance for lower duration segments. As indicated in Figure \ref{fig:Duration}, more than one third of the test segments have speech duration of less than 20 seconds. We have addressed this issue by proposing a short duration variability compensation method. The proposed method works by first extracting from each audio segment in the unlabelled development set, a partial excerpt of 10 seconds of speech material with random selection of the starting point (Figure \ref{fig:short}). Each audio file in the unlabelled development set, with the extracted audio segment will result in two 400-dimensional \emph{i}-vectors, one with at most 10 seconds of speech material. Considering each pair as one class, we computed a $400\times390$ LDA projection matrix to remove directions attributed to duration variability. Moreover, the projected \emph{i}-vectors are also subjected to a within-class covariance normalization (WCCN) using the same class labels.

\subsection{Language Normalization}
\label{sec:LN-LDA}
Language-source normalization is an effective technique for reducing language dependency in the state-of-the-art \emph{i}-vector/PLDA speaker recognition system \cite{mclaren2012language}. It can be implemented by extending SN-LDA \cite{mclaren2012source} in order to mitigate variations that separate languages. This can be accomplished by using the language label to identify different sources during training. Language Normalized-LDA (LN-LDA) utilizes a language-normalized within-speaker scatter matrix $\mathbf{\hat{S}}_{W}$ which is estimated as the variability not captured by the between-speaker scatter matrix,
\begin{equation}
    \hat{\mathbf{S}}_{W}=\mathbf{S}_{T} - \hat{\mathbf{S}}_{B},
    \label{eqn:SW}
\end{equation}
where $\mathbf{S}_{T}$ and $\hat{\mathbf{S}}_{B}$ are the total scatter and normalized between-speaker scatter matrices respectively, and are formulated as follows:
\begin{equation}
    {\mathbf{S}}_{T}=\sum\limits_{n=1}^{N}{\mathbf{w}_{n}{\mathbf{w}_{n}}^{T}},
\end{equation}
where $N$ is the total number of \emph{i}-vectors and
\begin{equation}       
\mathbf{\hat{S}}_{B}=\sum\limits_{l=1}^{L}{\sum\limits_{s=1}^{S_{l}}{n_{s}^{l}(\bar{\mathbf{w}}^{l}(s)-{\bar{\mathbf{w}}^{l}}){{(\bar{\mathbf{w}}^{l}(s)-{\bar{\mathbf{w}}^{l}})}^{T}}}},
\end{equation}
where $L$ is the number of languages in the training set, ${S}_{l}$ is the number of speakers in language $l$, $\bar{\mathbf{w}}^{l}(s)$ is the mean of $n_{s}^{l}$ \emph{i}-vectors from speaker $s$ and language $l$ and finally $\bar{\mathbf{w}}^{l}$ is the mean of all \emph{i}-vectors in language $l$. We applied a $390\times300$ SN-LDA projection matrix to reduce the \emph{i}-vector dimensions down to 300. 

\subsection{PLDA}
\label{sec:PLDA}
Probabilistic Linear Discriminant Analysis (PLDA) provides a powerful mechanism to distinguish between-speaker variability, separating sources which characterizes speaker information, from all other sources of undesired variability that characterize distortions. Since \emph{i}-vectors are assumed to be generated by some generative model, we can break it down into statistically independent speaker- and session-components with Gaussian distributions \cite{garcia2011analysis,kenny2010bayesian}. Although it has been shown that their distribution follow Student’s $t$ rather than Gaussian \cite{kenny2010bayesian} distributions, length normalizing the entire set of \emph{i}-vectors as a pre-processing step can approximately Gaussianize their distributions \cite{garcia2011analysis} and as a result improve the performance of Gaussian PLDA to that of heavy-tailed PLDA \cite{kenny2010bayesian}. A standard Gaussian PLDA assumes that an \emph{i}-vector $\mathbf{w}$, is modelled according to
\begin{equation}
    \mathbf{w}=\mathbf{m}+\mathbf{V}\mathbf{y}+\mathbf{\varepsilon}.
    \label{eq:plda}
\end{equation}
where, $\mathbf{m}$ is the mean of \emph{i}-vectors, the columns of matrix $\mathbf{V}$ contains the basis for the between-speaker subspace, the latent identity variable $\mathbf{y}\sim\mathcal{N}(\mathbf{0},\mathbf{I})$ denotes the speaker factor that represents the identity of the speaker and the residual $\mathbf{\varepsilon}$ which is normally distributed with zero mean and full covariance matrix $\mathbf{\Sigma}$, represents within-speaker variability. 

For each acoustic feature we have trained two PLDA models. The first out-domain PLDA ($\mathbf{V}_{out}$,$\mathbf{\Sigma}_{out}$) is trained using the training set presented in Table \ref{tab:data-nist}, and the second in-domain PLDA ($\mathbf{V}_{in}$,$\mathbf{\Sigma}_{in}$) was trained using the unlabelled development set. Our efforts to cluster the development set (e.g using the out-domain PLDA) was not very successful as it sounds that almost all of them are uttered by different speakers. Therefore, each \emph{i}-vector was considered to be uttered by one speaker. We also set the number of speaker factors to 200.

\subsection{Domain Adaptation}
\label{sec:domain}
Domain adaptation has gained considerable attention with the aim of compensating for cross-speech-source variability of in-domain and out-of-domain data. The framework presented in \cite{garcia2014unsupervised} for unsupervised adaptation of out-domain PLDA parameters resulted in better performance for in-domain data. Using in-domain and out-domain PLDA trained in Section \ref{sec:PLDA}, we interpolated their parameters as follow:
\begin{equation}
\begin{split}
    \mathbf{V}_{adapt}&=\alpha\mathbf{V}_{in}+(1-\alpha)\mathbf{V}_{out} \\
    \mathbf{\Sigma}_{adapt}&=\alpha\mathbf{\Sigma}_{in}+(1-\alpha)\mathbf{\Sigma}_{out}.
\end{split}
\label{eq:domainadapt}
\end{equation}
We chose $\alpha=0.10$ for making our submission.

\subsection{Score Computation and Normalization}
\label{sec:scorenorm}
For the one-segment enrolment condition, the speaker model is the length normalized \emph{i}-vector of that segment, however, for the three-segment enrolment condition, we simply used a length-normalized mean vector of the length-normalizated \emph{i}-vectors as the speaker model.
Each speaker model is tested against each test segment as in the trial list. For each two trial \emph{i}-vectors $\mathbf{w}_{1}$ and $\mathbf{w}_{2}$, the PLDA score is computed as 
\begin{equation}
    s=\mathbf{w}_{1}^{T}\mathbf{Q}\mathbf{w}_{1}+\mathbf{w}_{2}^{T}\mathbf{Q}\mathbf{w}_{2}+2\mathbf{w}_{1}^{T}\mathbf{P}\mathbf{w}_{2} + c,
\end{equation}
in which 
\begin{equation}
    \mathbf{Q}=\mathbf{S}_{T}^{-1} - (\mathbf{S}_{T}-\mathbf{S}_{B}\mathbf{S}_{T}^{-1}\mathbf{S}_{B})^{-1},
\end{equation}
\begin{equation}
    \mathbf{P}=\mathbf{S}_{T}^{-1}\mathbf{S}_{B}(\mathbf{S}_{T} - \mathbf{S}_{B}\mathbf{S}_{T}^{-1}\mathbf{S}_{B})^{-1}.
\end{equation}
and $\mathbf{S}_{B}=\mathbf{V}_{adapt}{\mathbf{V}_{adapt}}^{T}$ and $\mathbf{S}_{T}=\mathbf{S}_{B}+\mathbf{\Sigma}_{adapt}$.
It has been shown and proved in our experiments that score normalization can have a great impact on the performance of the recognition system. We used the symmetric s-norm proposed in \cite{kenny2010bayesian} which normalizes the score $s$ of the pair $(w_{1},w_{2})$ using the formula
\begin{equation}
    \hat{s}=\frac{s-\mu_{1}}{\sigma_{1}}-\frac{s-\mu_{2}}{\sigma_{2}}
\end{equation}
where the means $\mu_{1},\mu_{2}$ and standard deviations $\sigma_{1},\sigma_{2}$ are computed by matching $w_{1}$ and $w_{2}$ against the unlabelled set as the impostor speakers, respectively.

\subsection{Quality Measure Function}
\label{sec:QMF}
It has been shown that there is a dependency between the value of the $C_{det}^{min}$ threshold and the duration of both enrolment and test segments. Applying the quality measure function (QMF) \cite{novoselov2014stc} enabled us to compensate for the shift in the $C_{det}^{min}$ threshold due to the differences in speech duration. We conducted some experiments to estimate the dependency between the $C_{det}^{min}$ threshold shift on the duration of test segment and used the following QMF for PLDA verfication scores:
\begin{equation}
    QMF(t)=-0.2\sqrt{t}
\end{equation}
where $t$ is the duration of the test segment in seconds.

\subsection{Calibration}
\label{sec:calibrate}
In the literature, the performance of speaker recognition is usually reported in terms of calibrated-insensitive equal error rate (EER) or the minimum decision cost function ($C_{det}^{min}$). However, in real applications of speaker recognition there is a need to present recognition results in terms of calibrated log-likelihood-ratios. We have utilized the BOSARIS Toolkit \cite{brummer2011bosaris} for calibration of scores. $C_{det}^{min}$ provides an ideal reference value for judging calibration. If $C_{det}-C_{det}^{min}$ is minimized, then the system can be said to be well calibrated. 

The choice of target probability ($P_{tar}$) had a great impact on the performance of the calibration. However, we set $P_{tar}=0.0001$ for our primary submission which performed the best on the development set. For our secondary submission $P_{tar}=0.001$ was used. 

\begin{table*}[ht]
  \renewcommand{\arraystretch}{1}
  \setlength{\tabcolsep}{4pt}
  \vspace{-2mm}
  \caption{\label{tab:primary}\it{Performance comparison of the Intelligent Voice speaker recognition system with various analysis on the development protocol of NIST SRE 2016.}}
  \vspace{1mm}
  \centerline
  {
    \begin{tabular}{l l l c c c c c c}
      \toprule
      & & \multicolumn{3}{c}{\bfseries{Unequalized}} & & \multicolumn{3}{c}{\bfseries{Equalized}} \\
      \cmidrule{3-5} \cmidrule{7-9}
      \bf Acoustic Features & & EER &  $C_{det}^{min}$ & $C_{det}$ & & EER & $C_{det}^{min}$ & $C_{det}$ \\
      \midrule
      \bf Primary \\
      \rowgroup{MFCC} & & 16.49 & 0.6633 & 0.6754 & & 15.83 & 0.6650 & 0.6749   \\
      \rowgroup{PLP}  & & 17.87 & 0.6857 & 0.6977 & & 16.84   & 0.6914   & 0.6982 \\
      \rowgroup{Fusion}  & & \cg \bf 16.04   & \cg \bf 0.6012  & \cg \bf 0.6107 & & \cg \bf 14.93 & \cg \bf 0.6011 & \cg \bf 0.6267 \\
      \midrule
      \bf Scenario A \\
      \rowgroup{MFCC}  & & 16.82 & 0.6658 & 0.6794 & & 16.42 & 0.6890 & 0.7021   \\
      \rowgroup{PLP}   & & 16.98 & 0.6691 & 0.6881 & & 16.28   & 0.6903   & 0.7092 \\
      \rowgroup{ Fusion} & & 15.73   & 0.6153  & 0.6369 & & 15.12 & 0.6587 & 0.6964 \\
      \midrule
      \bf Scenario B \\
      \rowgroup{MFCC} & & 16.55 & 0.6735 & 0.6880 & & 16.10 & 0.6755 & 0.6945   \\
      \rowgroup{PLP}  & & 18.27 & 0.6938 & 0.7141 & & 16.97   & 0.7018   & 0.7299 \\
      \rowgroup{Fusion} & & 16.31   & 0.6075  & 0.6299 & & 14.70 & 0.6259 & 0.6482 \\
      \midrule
      \bf Scenario C \\
      \rowgroup{MFCC} & & 17.08 & 0.6767 & 0.6889 & & 16.77 & 0.6677 & 0.6927   \\
      \rowgroup{PLP}  & & 17.98 & 0.6857 & 0.6968 & & 17.21   & 0.7001   & 0.7192 \\
      \rowgroup{Fusion} & & 16.59   & 0.6176  & 0.6264 & & 15.70 & 0.6363 & 0.6680 \\
      \midrule
      \bf Scenario D \\
      \rowgroup{MFCC} & & 17.42 & 0.6694 & 0.6833 & & 16.54 & 0.6639 & 0.6820   \\
      \rowgroup{PLP}  & & 18.49 & 0.6851 & 0.7062 & & 17.46   & 0.6852   & 0.7054 \\
      \rowgroup{Fusion} & & 17.03   & 0.6171  & 0.6315 & & 15.73 & 0.6243 & 0.6410 \\
      \midrule
      \bf Scenario E \\
      \rowgroup{MFCC} & & 16.65 & 0.6976 & 0.7124 & & 16.24 & 0.6972 & 0.7122   \\
      \rowgroup{PLP}  & & 18.48 & 0.7182 & 0.7324 & & 17.49   & 0.7263   & 0.7480 \\
      \rowgroup{Fusion} & & 16.82   & 0.6343  & 0.6500 & & 15.52 & 0.6471 & 0.6737 \\
      \bottomrule
    \end{tabular}
  }
\vspace{-2mm}
\end{table*}

\section{Results and Discussion}
\label{sec:results}
In this section we present the results obtained on the protocol provided by NIST on the development set which is supposed to mirror that of evaluation set. The results are shown in Table \ref{tab:primary}. The first part of the table indicates the result obtained by the primary system. As can be seen, the fusion of MFCC and PLP (a simple sum of both MFCC and PLP scores) resulted in a relative improvement of almost 10\%, as compared to MFCC alone, in terms of both $C_{det}$ and $C_{det}^{min}$.
In order to quantify the contribution of the different system components we have defined different scenarios. In scenario A, we have analysed the effect of using LDA instead of NDA. As can be seen from the results, LDA outperforms NDA in the case of PLP, however, in fusion we can see that NDA resulted in better performance in terms of the primary metric. In scenario B, we analysed the effect of using the short-duration compensation technique proposed in Section \ref{sec:short}. Results indicate superior performance using this technique. In scenario C, we investigated the effects of language normalization on the performance of the system. If we replace LN-LDA with simple LDA, we can see performance degradation in MFCC as well as fusion, however, PLP seems not to be adversely affected. The effect of using QMF is also investigated in scenario D. Finally in scenario E, we can see the major improvement obtained through the use of the domain adaptation technique explained in Section \ref{sec:domain}.
For our secondary submission, we incorporated a disjoint portion of the labelled development set (10 out of 20 speakers) in either LN-LDA and in-domain PLDA training. We evaluated the system on almost 6k out of 24k trials from the other portion to avoid any over-fitting, particularly important for the domain adaptation technique. This resulted in a relative improvement of 11\% compared to the primary system in terms of the primary metric. However, the results can be misleading, since the recording condition may be the same for all speakers in the development set. 

\begin{table*}[t]
  \renewcommand{\arraystretch}{1}
  \setlength{\tabcolsep}{4pt}
  \vspace{-2mm}
  \caption{\label{tab:time}\it{CPU execution time and the amount of memory required to process a single trial.}}
  \vspace{1mm}
  \centerline
  {
    \begin{tabular}{l c c c c c c c}
      \toprule
       & segment dur & speech dur & stage & user time & system time & real time & memory(MB)  \\
      \midrule
      \multirow{3}{*}{Enrolment} &  \multirow{3}{*}{140s} & \multirow{3}{*}{60s} & features & 0.99s & 0.02s & 1.02s & 12 \\ 
      & & & SAD & 8.10s & 0.23s & 2.26s & 25.0  \\ 
      & & & \emph{i}-vectors & 27.47s & 2.20s & 7.83s & 4,014 \\
      \midrule
      \multirow{3}{*}{Test} &  \multirow{3}{*}{36s} & \multirow{3}{*}{25s} & features & 0.52s & 0.01s & 0.54s & 7.5 \\ 
      & & & SAD & 2.26s & 0.09s & 0.94s & 25.35  \\ 
      & & & \emph{i}-vectors & 26.02s & 2.25s & 7.9s & 4,013 \\
      \bottomrule
    \end{tabular}
  }
\vspace{-2mm}
\end{table*}

\section{Time Analysis}
\label{sec:timing}
This section reports on the CPU execution time (single threaded), and the amount of memory used to process a single trial, which includes the time for creating models from the enrolment data and the time needed for processing the test segments. The analysis was performed on an Intel(R) Xeon(R) CPU E5-2670 2.60GHz. The results are shown in Table \ref{tab:time}. We used the time command in Unix to report these results. The \emph{user time} is the actual CPU time used in executing the process (single thread). The \emph{real time} is the wall clock time (the elapsed time including time slices used by other processes and the time the process spends blocked). The \emph{system time} is also the amount of CPU time spent in the kernel within the process. We have also reported the memory allocated for each stage of execution. The most computationally intensive stage is the extraction of \emph{i}-vectors (both MFCC- and PLP-based \emph{i}-vectors), which also depends on the duration of the segments. For enrolment, we have reported the time required to extract a model from a segment with a duration of 140 seconds and speech duration of 60 seconds. The time and memory required for front-end processing are negligible compared to the \emph{i}-vector extraction stage, since they only include matrix operations. The time required for our SAD is also reported which increases linearly with the duration of segment. 

\section{Conclusions and Perspectives}
\label{sec:conclusion}
We have presented the Intelligent Voice speaker recognition system used for the NIST 2016 speaker recognition evaluation. Our system is based on a score fusion of MFCC- and PLP-based \emph{i}-vector/PLDA systems. We have described the main components of the system including, acoustic feature extraction, speech activity detection, \emph{i}-vector extraction as front-end processing, and language normalization, short-duration compensation, channel compensation and domain adaptation as back-end processing. For our future work, we intend to use the ALISP segmentation technique \cite{chollet1999toward} in order to extract meaningful acoustic units so as to train supervised GMM or DNN models. 

\bibliographystyle{IEEEbib}
\bibliography{references}

\end{document}